\documentclass{aa} 
\usepackage{epsfig,times} 
\usepackage{amssymb} 
\usepackage{txfonts} 
\usepackage{graphicx} 
\usepackage{natbib} 
\newcommand{\be}{\begin{equation}}
\newcommand{\ee}{\end{equation}}
\newcommand{\ba}{\begin{eqnarray}}
\newcommand{\ea}{\end{eqnarray}}

\begin{document} 
 
\title{Magneto--thermal evolution of neutron stars}
 
 
\author{J.A.~Pons\inst{1}, J.A.~Miralles\inst{1} \and U.~Geppert\inst{2}}
\institute{Departament de F\'{\i}sica Aplicada, Universitat d'Alacant,  
Ap. Correus 99, 03080 Alacant, Spain
\and 
German Aerospace Center, Institute for Space Systems, Rutherfordstr. 2, 12489 Berlin, Germany}
\date{Received...../ Accepted.....} 
 
\abstract 
{The presence of magnetic fields in the crust of neutron stars causes a
non-spherically symmetric temperature distribution. The strong temperature
dependence of the magnetic diffusivity and thermal conductivity,
together with the heat generated by magnetic dissipation,
couple the magnetic and thermal evolution of NSs, that
cannot be formulated as separated one--dimensional problems.
}
{We study the mutual influence of thermal and magnetic evolution in
a neutron star's crust in axial symmetry. Taking into account realistic 
microphysical inputs, we find the heat released by Joule effect consistent with the
circulation of currents in the crust, and we incorporate its effects
in 2--dimensional cooling calculations. }
{We solve the induction equation numerically
using a hybrid method (spectral in angles, but a finite--differences scheme 
in the radial direction), coupled to the thermal diffusion equation.
To improve the boundary conditions, we also revisit the envelope stationary solutions updating 
the well known $T_{\rm b}-T_{\rm s}$--relations to include the effect of 2--D heat transfer 
calculations and new microphysical inputs.
} 
{We present the first long term 2--dimensional simulations of the coupled magneto-thermal
evolution of neutron stars. This substantially improves previous works in which
a very crude approximation in at least one of the parts (thermal or magnetic diffusion)
has been adopted. Our results show that the feedback between Joule heating and
magnetic diffusion is strong, resulting in a faster dissipation of the stronger fields
during the first $10^5-10^6$ years of a NS's life. As a consequence, 
all neutron stars born with fields larger than a critical value ($> 5 \times 10^{13} G$)
reach similar field strengths ($\approx 2-3 \times 10^{13} G$) at late times.
Irrespectively of the initial magnetic field strength, after $10^6$ years the 
temperature becomes so low that the magnetic 
diffusion timescale becomes longer than the typical ages of radio--pulsars, 
thus resulting in apparently no dissipation of the field in old NS. We also
confirm the strong correlation between the magnetic field and the surface temperature 
of relatively young NSs discussed in preliminary works. The effective temperature of
models with strong internal toroidal components are systematically higher than those
of models with purely poloidal fields, due to the additional energy reservoir stored
in the toroidal field that is gradually released as the field dissipates.
} 
{} 
 
\keywords{Stars: neutron - Stars: evolution - Stars: magnetic fields}
\titlerunning{Magneto-thermal evolution of NSs} 
\authorrunning{J.A. Pons, J.A. Miralles, \& U. Geppert} 
 
\maketitle 
 
\section{Introduction} 

The neutron star (NS) magnetic field (MF) maintained by electric currents circulating in 
the crust modifies the crustal temperature distribution by means of two mechanisms.
The first one is due to the anisotropy of thermal conductivity in presence of a 
strong MF and causes important changes on how heat flux flows from the 
core through the crust and envelope up to the surface. The second mechanism is 
the generation of heat due to MF dissipation which results in a non 
spherically symmetric allocation of heating sources.
Therefore, the magnetic diffusivity, besides its tensorial character due to the
presence of the MF (which eventually results in the Hall drift of the 
field), becomes a non-spherically symmetric quantity.
Under these conditions, an initially purely dipolar field evolves very quickly in a magnetized
NS to generate complex structures including toroidal fields and/or higher order 
multipoles. Different studies on magneto-hydrostatic equilibrium configurations indicate
that stable configurations require both toroidal and poloidal components
(see e.g. \cite{Reis2008} and references therein).

The complexity of the problem has limited previous works
to partial studies of the complete problem.
The anisotropic heat flux and its consequences for the - in principle observable - 
surface temperature ($T_{\rm s}$) distribution has been considered by 
\cite{Geppert2004,Geppert2006} and \cite{Azorin2006a,Azorin2006b}, but for
fixed, prescribed MF. 
The long--term effect of Joule heating has also been recently discussed by 
\cite{Aguilera2008a} or \cite{UK2008}
(see also references therein), although in a very crude approach. 
In \cite{Aguilera2008b} a full 2-D cooling code
has been used to describe the temperature evolution, but the Joule heating rate is estimated
by an analytical approximation and considered to be uniform. In \cite{UK2008}
the former results of \cite{Miralles1998} are revisited in the context
of high field radio--pulsars, but with the simplified approach of considering the 
diffusion of a one--mode (dipolar) poloidal field, without actually performing consistent 
simulations including the temperature evolution.
On the other hand, the evolution of the MF including 
the influence of the Hall term at early times has been studied
in \cite{PonsGeppert2007}. The Hall drift causes a somewhat faster dissipation for
strong initial fields due to the reorganization of the field in smaller scales.
In this latter work the temporal evolution of the temperature
has been prescribed according to a generally accepted cooling law 
and assumed to be the same in the whole crust. The effect of ambipolar
diffusion in the liquid core can also be relevant, as recently studied
for example in \cite{Hoyos2008}.

One of the reasons that leads to complex MF geometries in the 
crust is due to the non-spherically symmetric magnetic 
diffusivity. It is effective even if the Hall drift is negligible. 
Since the temperature within the crust is no longer uniform,
the magnetic diffusivity in this approach 
becomes a function of the radial and of the polar coordinate, thus 
rendering the one--mode approximation inappropriate.
Up to now, the effect of the dependence of the magnetic diffusivity on the polar
angle $\theta$ (if axial symmetry is assumed) has not been considered.
In some studies about cooling of NSs, the source term in the thermal diffusion equation 
includes an angular--averaged Joule heating rate, 
where the heat production is uniform in spherical shells at a given radius $r$ 
\citep{Page2000,Aguilera2008a,UK2008}. These are not fully consistent calculations,
because the currents can locally be very intense and release heat in a highly non-spherically
symmetric
way. Therefore, it is important to attack the problem of the full magneto-thermal
evolution of NSs, by solving simultaneously the induction equation and the heat transfer
equation taking into account the anisotropy of the thermal conductivity and the electrical
resistivity tensors in a consistent way. This is the main goal of this paper, in which we
solve this problem (in axial symmetry) for the first time.
The aim of this study is to provide a self-consistent model for the coupled
evolution of the MF and temperature in axially symmetric NSs, in general relativity,
and including the effect of angular variations of temperature in the magnetic 
diffusivity and the thermal conductivity. 
We also revisit the envelope stationary solutions updating 
the well known $T_{\rm b}-T_{\rm s}$--relations to include the effect of 2--D heat transfer 
calculations and new advances in microphysical ingredients (phonons, ion--ion interactions). 
We provide new fits to solutions of equilibrium magnetic envelope that can be used 
as boundary conditions in multidimensional cooling simulations.

The paper is organized as follows. In the next Section we derive and present the 
basic equations together with the boundary conditions. Next we describe the input 
microphysics and discuss the effect of superfluid neutrons on the heat flux 
anisotropy. The following Section is devoted to the presentation of the 
magneto--thermal evolution. Finally we discuss the results by comparing them with 
observable consequences.\\

\section{Basic Equations}
The thermal evolution of the NS is calculated using the code described
in \cite{Aguilera2008b}. For the evolution of the MF
we basically follow the formalism already used in \cite{PonsGeppert2007}.
We address the interested reader to these references for more details about
the cooling code and the MF evolution code that we have merged 
and coupled to perform the simulations presented later in this paper. 
In the next subsections we just summarize the basic equations
that are solved with the purpose of introducing notation and
for the sake of completeness.

\subsection{Heat Transfer Equation}

Assuming that deformations with respect to the spherically symmetric case due 
to rotation, MF, and temperature distribution do not affect 
the metric in the interior of a NS, we use the standard static metric 
\be
ds^2= -c^2 e^{2\nu(r)}dt^2+ e^{2\lambda(r)}dr^2 + r^2 d\Omega^2~,
\label{metric}
\ee
where $\nu(r)$ is the relativistic redshift correction and $\lambda(r)$
is the length correction factor, $ e^{-2\lambda(r)} = 1 - 2 G M(r)/c^2 r$.
Using this background metric,
the thermal evolution of a NS can be described by the energy balance 
equation 
\be 
c_{\rm v} e^{\nu} \frac{\partial T}{\partial t} + \vec{\nabla} \cdot  
(e^{2\nu } \vec{F})= 
e^{2 \nu} (-Q_{\nu} + Q_h)  
\label{eneq} 
\ee 
where $c_{\rm v}$ is the specific heat per unit volume and $Q_{\nu}$ is the 
energy loss by neutrino emission while $Q_h$ stands for the Joule heating rate. 
In this study we neglect other heating mechanisms than Joule heating, such
as accretion or frictional heating at the core crust interface. 
In the diffusion limit, the heat flux is simply 
\be 
\vec{F} = -e^{-\nu} \hat{\kappa} \cdot \vec{\nabla} (e^{\nu} T) 
\label{fluxeq} 
\ee 
where $\hat \kappa$ is the thermal conductivity tensor that depends on both
coordinates $(r,\theta)$ through its dependence on temperature and MF.
Note that the differential operator $\nabla$ associated to the spatial metric 
$e^{2\lambda(r)}dr^2 + r^2 d\Omega^2$ must include the corresponding metric factors (i.e. 
$e^{-\lambda(r)} \frac{\partial}{\partial r}$ for
the radial component). An explicit representation in terms of the metric
scale factors can be found in \cite{Radler2001}. We solve Eqs. (\ref{eneq}) 
and (\ref{fluxeq}) numerically following the same scheme as described 
in \cite{Geppert2004,Azorin2006a}.

\subsection{The Induction Equation}
The Maxwell equations (in the Gaussian system) relative to an observer at rest (Eulerian observer) 
for the static metric given by Eq. (\ref{metric}) are
\begin{eqnarray}
\label{ME}
\nabla\cdot \vec{E}&=&4\pi\,\rho_e \nonumber \\
\frac{1}{c}\frac{\partial \vec{E}}{\partial t}&=&\vec{\nabla}\times(e^\nu\,\vec{B})-\frac{4\pi}{c}\, e^\nu\, \vec{j} \nonumber \\
\vec{\nabla}\cdot \vec{B}&=&0 \nonumber \\
\frac{1}{c}\frac{\partial \vec{B}}{\partial t}&=&-\vec{\nabla}\times(e^\nu\,\vec{E})
\end{eqnarray}
where again the differential operators are defined according to the spatial coordinate system,
$\vec{E}$ and 
$\vec{B}$ are the electric and magnetic field measured by the Eulerian observer, as well as 
$\rho_e$ and $\vec{j}$ that represent the electric charge and current density, respectively.

In the comoving frame, which corresponds to the Eulerian frame since matter is at rest, 
electric field and density current are related through Ohm's law. This 
law establishes a proportionality between the current and the electric field which can be written 
as $\vec{j}=\sigma \vec{E}$, $\sigma$ being the electric conductivity. In the presence of strong
MFs, $\sigma$ becomes a tensor and Ohm's law must be
written as $\vec{j}=\hat{\sigma} \vec{E}$ or equivalently $\vec{E}=\hat{\cal R}\vec{j}$, where
$\hat{\cal R}\equiv \hat{\sigma}^{-1}$ is the resistivity tensor. 
This tensor can be decomposed in symmetric and antisymmetric parts \citep{LandauEDCM}. 
The symmetric part, for non-quantizing fields is isotropic and determined by the scalar 
$\frac{1}{\sigma_\parallel}$, where $\sigma_\parallel$ is the electric conductivity 
in the direction of the MF. The antisymmetric part of the resistivity tensor can
be represented by a vector proportional to $\vec{B}$.
The electric field $\vec{E}$ is then written in the form \citep{ziman}
\begin{equation}
\label{Ohmlaw}
\vec{E}=\frac{1}{\sigma_\parallel}\vec{j}+\frac{1}{e n_e c}\vec{B}\times\vec{j}
\end{equation}
where $n_e$ is the electron number density and $e$ is the elementary charge.

If we neglect the displacement current term in the Amp\`ere-Maxwell equation 
(second equation in \ref{ME}), and make use of Eq. (\ref{Ohmlaw}), the Faraday induction 
equation (fourth equation in \ref{ME}) can be written as
\begin{equation}
\label{Hallind}
\frac{\partial \vec{B}}{\partial t}= -\vec{\nabla}\times\left\{\eta\vec{\nabla}\times (e^{\nu}\vec{B})+
\frac{c}{4\pi e n_e}\left[\vec{\nabla}\times(e^{\nu}\vec{B})\right]\times
\vec{B}\right\}
\end{equation}
where we have introduced the magnetic diffusivity
$\eta\equiv\frac{c^2}{4\pi\sigma_\parallel}$.

There are two main differences from the induction equation employed in \cite{PonsGeppert2007}:
the relativistic factors are included and the magnetic diffusivity is not assumed
to be spherically symmetric.

Since both $n_e$ and the metric factor $e^\nu$ only depend on the radial coordinate 
(it is well justified to neglect the structural deformations induced by MFs) 
the non--linear Hall term
can be treated in the same way as done by \cite{PonsGeppert2007}. 
However, the magnetic diffusivity through its temperature dependence
also depends on the polar angle, $\eta=\eta(r,\theta)$, thus requiring an extension of 
the previous formalism.
The influence of the Hall term at early times has been studied
in \cite{PonsGeppert2007}, resulting in a somewhat faster dissipation for
strong initial fields due to the reorganization of the field in smaller scales.
The numerical treatment of the non-linear term is complex and very 
computationally--limited. 
In this paper our goal is to perform coupled (magneto-thermal) long term simulations,
thus from now on we focus on the linear part of the induction equation. 

The linear part of the induction equation (\ref{Hallind})  reads
\be 
\frac{\partial\vec B}{\partial t}= - \nabla \times\left[\eta 
~\nabla \times (e^{\nu} \vec{B}) \right].
\label{Hallind_lin} 
\ee
We decompose the MF into its poloidal and toroidal part
\citep{Radler2001}
\be
\vec{B} = \vec{B}_{\rm pol} + \vec{B}_{\rm tor}\,,
\ee
and use their representation in terms of two scalar functions $\Phi(r,\theta)$ 
and $\Psi(r,\theta)$:
\be
\vec{B}_{\rm pol}= -\vec{\nabla} \times \left( \vec{r}\times\vec{\nabla}\,\Phi \right),~~~
\vec{B}_{\rm tor}= -\vec{r}\times\vec{\nabla}\,\Psi\,.
\label{MF_scalars}
\ee
We expand now the functions $\Phi$, $\Psi$, and $\eta$ in a series of spherical 
harmonics (in the axisymmetric case) as follows:
\ba 
\Phi &=& \frac{1}{r} \sum_{n=1}^{\infty} \Phi_{n}(r,t) Y_{n}(\theta)~, 
\nonumber \\ 
\Psi &=& \frac{1}{r} \sum_{n=1}^{\infty} \Psi_{n}(r,t) Y_{n}(\theta)~,
\nonumber \\
\eta &=& \sum_{n=0}^{\infty} \eta_{n}(r,t) Y_{n}(\theta)~,
\label{expans} 
\ea 
where $Y_{n}$ is the spherical harmonic $Y_{n}^{m}$ for $m=0$.

The poloidal and toroidal parts of the MF can be written as
\ba
\vec{B}_{\rm pol} &=& \frac{1}{r^2} \sum_{n} n(n+1) \Phi_{n} Y_{n}~ \vec{e}_{r}
+ \frac{e^{-\lambda}}{r} \sum_{n} \frac{\partial \Phi_{n}}{\partial r}
\frac{d Y_{n}}{d \theta}~\vec{e}_{\theta}~,
\nonumber \\
\vec{B}_{\rm tor} &=& - \left( \frac{1}{r} \sum_{n} \Psi_{n}
\frac{d Y_{n}}{d \theta} \right)~\vec{e}_{\phi}~,
\label{bpoltor}
\ea
where $\vec{e}_{r}, \vec{e}_{\theta}, \vec{e}_{\phi}$ are unitary vectors associated to
the spatial metric.
Inserting this expressions into Eq.~\ref{Hallind_lin} and using the quantities
$I^{n}_{k'k}$ and $I^{(1)}$ defined in Geppert \& Wiebicke (1991, Eqs. 59 and 60)
which are related to the Clebsch--Gordan coefficients,
we arrive at the following evolution equations for the poloidal and toroidal scalar functions:
\ba
\frac{\partial \Phi_n}{\partial t} &=& e^{\nu}
\sum_{k,k'}  \eta_{k'} ~ \left(I^{n}_{k'k}-I^{(1)}\right)
\times
\nonumber \\
&& \times
\left[ e^{-2\lambda} \frac{\partial^2 \Phi_k}{\partial r^2} 
+  e^{-2\lambda} \left( \frac{d \nu}{d r} - \frac{d \lambda}{d r} \right) 
\frac{\partial \Phi_k}{\partial r}  
-\frac{k(k+1)}{r^2} \Phi_k\right]
\label{Phi_lin_diff}
\ea
and
\ba
\frac{\partial \Psi_n}{\partial t}=\sum_{k,k'}
&& \left\{ \left(I^{n}_{k'k}-I^{(1)}\right)
e^{-\lambda} \frac{\partial}{\partial r} 
\left( \eta_{k'}  e^{-\lambda} \frac{\partial  (e^{\nu} \Psi_k)}{\partial r}\right)
\right.
\nonumber \\
&& \left.
- \eta_{k'}~ I^{n}_{k'k}\frac{k(k+1)}{r^2} e^{\nu} \Psi_k \right\} ~.
\label{Psi_diff}
\ea
If $\eta=\eta(r,t)$ only $k'=0$ contributes and, taking the 
non-relativistic limit ($e^{\nu}=e^{\lambda}=1$), the purely diffusive
Eqs. 8 of \cite{PonsGeppert2007} with $D_{nm}=C_{nm}=0$
are recovered. Note that no coupling between poloidal and toroidal component
appears, since this coupling can only be obtained by including the Hall drift term.

\subsection{Microphysics}

The microphysical ingredients that enter in the heat transport and
induction equation are the specific heat, the thermal conductivity,
the neutrino emissivity and the magnetic diffusivity.
In the solid crust,
the dominant contribution to the specific heat is that from electrons and ions, while
electrons, lattice phonons and collective modes of superfluid neutrons contribute 
to the thermal conductivity. We refer the reader to  
a detailed description of the employed equations of state, 
the thermal conductivity, the specific heat, and the neutrino emissivities, 
given in section 4 of \cite{Aguilera2008b}.
In this paper we assume the {\it minimal cooling scenario} 
\citep{Page2004} which includes neutrino emission from the Cooper pair breaking 
and formation process, but we do not take into account any 
direct Urca process, due either to nucleons or to exotic matter
(hyperons, Bose condensates or deconfined quarks).

In addition, it has been shown by \cite{Chugunov2007}  that the ionic contribution 
to the total thermal conductivity 
is negligible in the crust but it may play a role for low temperatures in the envelope. 
Very recently \citep{sphs} studied the possible effects of 
collective modes of superfluid neutrons.  
They found that this process may dominate the thermal conductivity in the inner
crust when its temperature is $\la 10^7$K. For such relatively old NSs, heat
transport by superfluid neutrons counteracts the anisotropy in the electron
conductivity caused by a strong crustal field and, eventually, 
turns the inner crust isothermal.
In this work we have updated our microphysics and we have also 
included this two new contributions to the thermal conductivity.

The only relevant contributions to the electrical resistivity are electron--phonon 
and electron--impurity collisions \citep{FI1976}. 
While the efficiency of electron--phonon collisions strongly depends 
on the crustal temperature, the electron--impurity scattering is much less sensitive to it. 
However, both processes may be strongly affected by the presence of a strong
MF which suppresses the conductivity components perpendicular to the field lines 
\citep{Canuto1970,Itoh1975}. As in \cite{PonsGeppert2007}, we calculate the
electrical resistivity by using the electron relaxation time provided by A. Potekhin's
public code \footnote{\tt  www.ioffe.rssi.ru/astro/conduct/condmag.html}.
We have used an impurity concentration parameter of $0.1$
(the definition of this parameter and a discussion about how it affects electronic
transport can be found in section 5.1.1 of \cite{Azorin2006a}).
\cite{Jones1999} has shown that disorder in the inner crust
could result in an impurity parameter $\gtrsim 10$, which leads to larger electrical 
resistivity and enhanced ohmic decay. However, this effect becomes important
only for sufficiently cool NSs. 

\subsection{Joule Heating} 
Joule heating couples the thermal and magnetic evolution by contributing
to the source term $Q_h$ in Eq.~\ref{eneq} which, for large fields, can result
in a higher temperature and therefore a larger magnetic diffusivity. These feedback
may lead to a faster dissipation of the MF if it is strong enough to really
alter the temperature of the crust.
In this paper Joule heating is taken consistently into account 
in a cooling simulation for the first time.
As the MF evolves in time, we compute
at each point of the computational grid 
the local values of the electrical current density, which is simply
\ba
\vec{j}(r,\theta,t) = e^{-\nu} \frac{c}{4 \pi} \nabla \times (e^{\nu} \vec{B})~.
\ea
The corresponding components of the current density are
\ba
\frac{4\pi}{c} \vec{j} &=& \frac{1}{r^2} \sum_{n} n(n+1) \Psi_{n} Y_{n} ~\vec{e}_{r}
\nonumber \\
&+& \frac{e^{-\lambda-\nu}}{r} \sum_{n} \frac{\partial (e^{\nu} \Psi_{n})}{\partial r}
\frac{d Y_{n}}{d \theta}~\vec{e}_{\theta}
\nonumber \\
&+& \frac{1}{r} \sum_{n} \left[ e^{-2\lambda}
\frac{\partial^2 \Phi_{n}}{\partial r^2} 
+ \frac{2 G M(r)}{c^2 r^2} \frac{\partial \Phi_{n}}{\partial r}
- \frac{n(n+1)}{r^2}\Phi_{n}
\right] \frac{d Y_{n}}{d \theta} ~\vec{e}_{\phi}.
\nonumber \\
\label{currents}
\ea

We evaluate the heating source term by
\be
Q_h = \frac{\vec{j}^2}{\sigma_\parallel}~,
\label{Joule_heat_rate}
\ee
where $Q_h$ is the energy per unit time and unit volume measured by the Eulerian observer. 
The magnetic energy balance equation is
\be
\frac{\partial}{\partial t} \left(e^{\nu}\frac{B^2}{8 \pi}\right) 
= - e^{2\nu}\frac{\vec{j}^2}{\sigma_\parallel} -
\nabla \cdot \left[ e^{2\nu}\frac{1}{4 \pi} \vec{E}  \times \vec{B} \right]~,
\label{econs}
\ee
which is easily interpreted: if we integrate over the whole volume, 
the magnetic energy losses are due to Joule heating and Poynting flux through
the boundaries. This latter term vanishes since we do not consider
the possibility of having electromagnetic waves (we neglect 
displacement currents).

\subsection{Magnetic boundary conditions}

Since we restrict ourself to MF configurations confined to the crust, 
the inner boundary conditions are determined by the requirement that 
the normal component of the MF and the tangential components of the electric field 
has to vanish at $r=R_c$. This is a consequence of the assumption 
that the core is in a superconducting state and the Meissner--Ochsenfeld effect 
prevents the MF to penetrate. Therefore we apply the following boundary conditions
at $r = R_c$
\ba 
\Phi_n &=& 0
\\
e^{-\lambda}\frac{\partial \Psi_n}{\partial r}&=&-\frac{\omega_{\rm B}\tau}{r^2} 
\sum_{k,k'} I^{(2)} \Psi_k\Psi_{k'}\;\;.
\label{IBC} 
\ea
A detailed derivation is given in \cite{PonsGeppert2007}.
Note that in the limit of vanishing Hall--drift this 
reduces to $\frac{\partial \Psi_n}{\partial r}=0$.

For the outer BC we require all components of the MF to be
continuous across $r=R_{\rm{NS}}$ to match the relativistic vacuum solution. Hence,
let us first consider the stationary solution for the outer space. In the absence
of external currents, the toroidal component of the MF must vanish, and each
multipole of the poloidal field must satisfy (in the stationary case):
\ba
(1-z) \frac{\partial^2 \Phi_n}{\partial r^2} 
+ \frac{z}{r}\frac{\partial \Phi_n}{\partial r}  
-\frac{n(n+1)}{r^2} \Phi_k = 0
\label{ode2}
\ea
where $z=2 G M/c^2 r$, which corresponds to the compactness parameter
at $r=R_{\rm NS}$.

This second order differential equation has analytical solutions for each
value of $n$, although they cannot be written in a closed analytic
form valid for any $n$. For example, for $n=1$ and $n=2$ we have
\be
\Phi_1 = C_1 r^2 \left[ \ln(1-z) 
+ z + \frac{z^2}{2}\right] + C_2 r^2
\ee
\be
\Phi_2 = C_1 r^3 \left[ (4-3z) \ln(1-z) 
+ 4z - {z^2} - \frac{z^3}{6}\right] + C_2 r^3 (4-3z)
\ee
where $C_1$ and $C_2$ are arbitrary integration constants that must be fixed according
to the value of the magnetic multipole moments. Regularity of the external solution 
at $r=\infty$ requires $C_2=0$.
In general, the family of solutions of Eq. \ref{ode2} for any value of $n$
can also be expressed in terms of generalized hypergeometric functions ($F([],[],z)$), 
also known as Barnes's extended hypergeometric functions, as follows:
\ba
\Phi_n &=&  C_1 ~r^{-n} ~F([n,n+2], [2+2n], z)
\nonumber \\
&+& C_2 ~r^{n+1} ~ F([1-n,-1-n], [-2n], z)~. 
\label{outerPhi}
\ea
Note that regularity at $r=\infty$ requires again $C_2=0$.

For practical reasons, we write the outer boundary conditions
at $r=R_{\rm NS}$ as 
\ba
\frac{\partial \Phi_{n}}{\partial r} &=& -\frac{n}{R_{\rm NS}} f_n \Phi_n
\\
\Psi_{n} &=&0 
\label{OBC} 
\ea
where $f_n$ is a relativistic factor that only depends on the compactness ratio
$z(R_{\rm NS})$ (in the Newtonian limit $f_n=1$) and can be evaluated numerically
or with the help of any algebraic manipulator using the form given in
Eq. (\ref{outerPhi}). \cite{Radler2001} use an alternative form based on
the expansion of the vacuum solutions in a series of powers of $1/r$.


\section{Thermal boundary conditions: blanketing envelopes revisited}

\begin{figure}[th]
\centering
\includegraphics[width=0.45\textwidth]{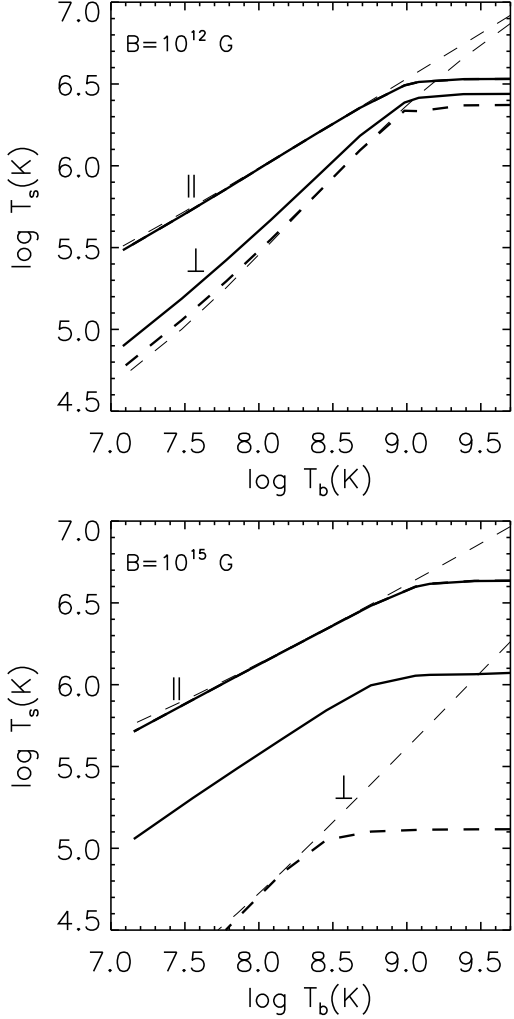}
\caption{Surface temperature as a function of temperature at the neutron drip
point for two models of NS envelopes with different MF strengths.
The thin dashed lines are the analytical fits from  \citep{Potekhin2001} for
MF parallel ($\parallel$) and perpendicular ($\perp$) to the normal direction to
the surface.  The thick
dashes are our results for the $\perp$ case, and the solid lines
are our full 2D results for a dipolar field (we show $T_s$ at the pole and at the equator). 
Our results for a purely radial
field cannot be distinguished from the upper solid line ($T_s$ at the pole for the
dipolar MF).
Numerical results always include the effect
of neutrino emission in the outer crust and envelope.
}
\label{fig1}
\end{figure} 

The very different thermal relaxation timescales of the envelope and the crust
of NSs makes computationally expensive 
any attempt to perform cooling simulations
in a numerical grid that includes both regions simultaneously.
Since radiative equilibrium is established in the low density region much faster
than the crust evolves, the usual approach is to use 
results of stationary, plane-parallel, envelope models to obtain a phenomenological fit 
that relates the temperature at the bottom of the envelope $T_{\rm b}$, with the surface 
temperature $T_{\rm s}$. This $T_{\rm s}=T_{\rm s}(T_{\rm b})$  phenomenological function 
is used to implement boundary 
conditions, because it allows to calculate the surface flux for a given temperature at 
the base of the envelope.
$T_{\rm b}$ is generally chosen to correspond to some density between the neutron drip point
$\rho \approx 3\times 10^{11}$g cm$^{-3}$ and $\rho=10^{10}$g cm$^{-3}$.
Examples of such models of magnetized envelopes have been constructed by \cite{Potekhin2001}
and later upgraded in \cite{Potekhin2007} to include the effect of the neutrino 
emissivity in the outer crust.
They derived an analytic form of the $T_{\rm b}-T_{\rm s}$ relation that reads
\be 
   T_\mathrm{s}(B,\varphi,g,T_{\rm b})\approx 
   T_\mathrm{s}^{(0)}(g,T_{\rm b}) \,\mathcal{X}(B,\varphi,T_{\rm b}) ,  
\label{PCY-iron} 
\ee 
where
\begin{equation} 
   T_\mathrm{s}^{(0)} \approx  10^6 \, 
   g_{14}^{1/4}\left[(7\zeta)^{2.25}+(\zeta/3)^{1.25}\right]^{1/4}~~{\rm K}, 
\end{equation} 
and 
$\zeta\equiv 0.1 T_{b,8} -0.001\,g_{14}^{1/4}\,\sqrt{0.7 \, T_{b,8}}$.
Here $g_{14}$ is the surface gravity in units of $10^{14}$ cm~s$^{-2}$, 
$T_{b,8}$ is $T_{\rm b}$ in $10^8$~K, and  $T_{s,6}$ is $T_{\rm s}$ in $10^6$~K.

The function $\mathcal{X}$ has been fitted by  
decomposing into transversal and longitudinal parts as 
\ba 
 \mathcal{X}(B,\varphi,T_{\rm b}) &=&  
      \big[ \,\mathcal{X}_\|^{9/2}(B,T_{\rm b})  \cos^2\varphi
   + \,\mathcal{X}_\perp^{9/2}(B,T_{\rm b}) \sin^2\varphi \big]^{2/9}~, 
 \label{fit3} 
\ea 
where $\varphi$ is the angle between the MF and the normal to the surface.
\cite{Potekhin2001} give the following fits for $\mathcal{X}_\perp$
and $\mathcal{X}_\|$
\ba
\mathcal{X}_\parallel(B,T_{\rm b}) &=& 1 + 0.0492 B_{12}^{0.292} T_{b,9}^{0.240}
\\
\mathcal{X}_\perp(B,T_{\rm b}) &=& \frac{\sqrt{1 + 0.1076 B_{12} (0.03+T_{b,9})^{-0.559}}}
{\left[1 + 0.819 B_{12}/ (0.03+T_{b,9})\right]^{0.6463}}
\label{fitsold}
\ea
where $B_{12}=B$ in units of $10^{12}$ G.
These fits are valid for $B < 10^{16}$~G and  $10^7 \mbox{ K} \leq T_{\rm b} \leq 10^{9.5}$~K.

It must be reminded that the above $T_{\rm b}-T_{\rm s}$ relation is based on a 
plane--parallel approximation. When this approach is applied to a spherical star, 
meridional heat fluxes in the envelope are not allowed and, therefore, this approximation 
may be inaccurate 
when these fluxes compete with the purely radial ones. In addition, 
\cite{Chugunov2007} found that the contribution of ions or phonons to the thermal
conductivity of the envelope can reduce the anisotropy of heat conduction. Therefore,
we have revisited the magnetized envelope problem with two motivations: upgrading the
microphysical inputs (thermal conductivity) and assessing on the accuracy of the
plane-parallel approximation. 

To avoid solving the hydrostatic equilibrium equations in two dimensions, we have
build a spherically symmetric iron envelope model with a zero temperature equation 
of state.  With this fixed background, we have calculated stationary solutions of 
the heat transport equation in 2D, with a given MF geometry (dipole solution).
At first glance, this may seem inaccurate, since finite temperature
effects may be relevant at low density but, as explained in \cite{Gudmundsson1983},
the main regulator of the  $T_{\rm b}-T_{\rm s}$--relation  is the sensitivity strip where the
opacity is maximum. This strip marks the transition from electronic heat transfer
to the radiation dominated one. It lies at relatively high densities except for very
low temperatures. Hence, structural changes at low density do not affect the
solution, even for extreme cases such as a condensed surface \citep{Potekhin2007}.
To test the validity of this assumption and to compare with previous works, 
we have taken the crustal MF to be either purely
radial (labelled $\parallel$, parallel to the normal to the surface) 
or purely tangential to the surface (labelled $\perp$).
In this two limiting cases, the heat flux is purely radial, and we have reproduced
within a 5\% the $T_{\rm b}-T_{\rm s}$ relation given by \cite{Potekhin2001} if
neutrino emissivities are not considered. The minor
differences are probably due to the different EOS or NS model (i.e. our neutron
drip point is at $\rho=3.5 \times 10^{11}$ g/cm$^3$, while theirs is at
$\rho=4 \times 10^{11}$. Our results for purely parallel or purely tangential
fields including the neutrino emissivity in the outer crust are shown with 
thick dashes in Fig. \ref{fig1}. The analytical fit from
\citep{Potekhin2001}, without the effect of neutrino emissivity,  is plotted with thin 
dashes. The agreement is very good in those cases where neutrino emission is not 
important (low temperature).
Moreover, our results for radial and tangential MF are directly comparable 
to Figs. 4 and 5 of \cite{Potekhin2007}, in which neutrino emissivity effects are 
considered.  The very good agreement with this work indicates that the use of a zero
temperature EOS for solving the hydrostatic equilibrium equation is a valid 
approximation. 
The same comment applies to the effects
of strong MF  in the low density EOS, as long as they are only noticeable at densities
where radiation dominates the heat transport. 
Finally, in order to check the effect of using an EOS in which the
thermal contribution to the pressure is neglected to obtain the mechanical 
structure of the envelope, we have also recalculated some of the 
models using new envelope structures obtained by solving the hydrostatic equilibrium
equations for a finite temperature 
EOS and assuming the temperature profile of the previous models. We have found the same 
surface temperatures in both models (within a 1\%), further confirming the validity of 
our approach.

\begin{figure}[th]
\centering
\includegraphics[width=0.45\textwidth]{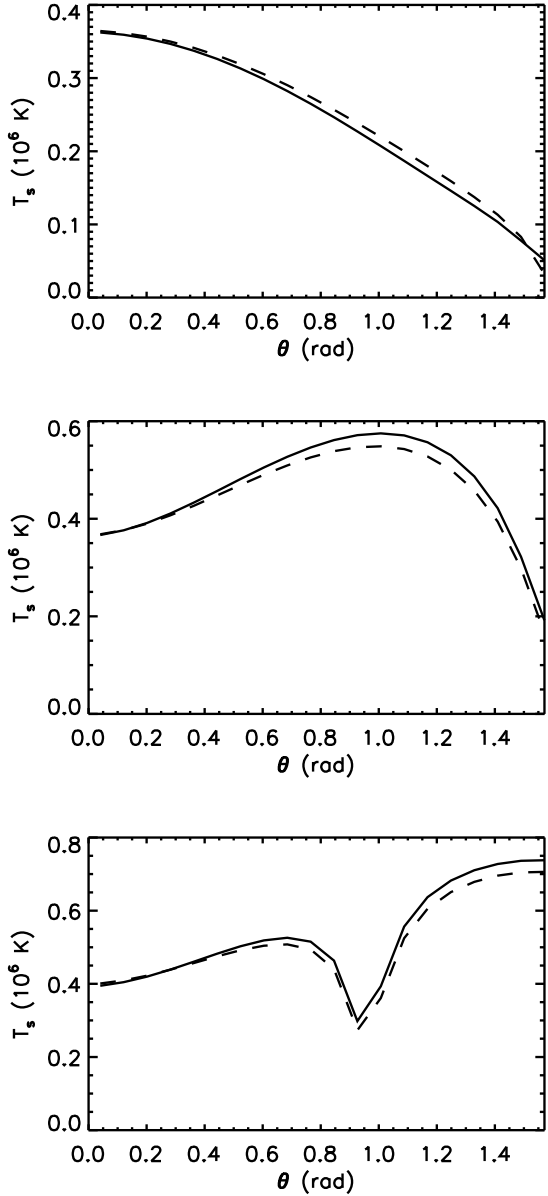}
\caption{Stationary 2D solutions of magnetized envelopes:
surface temperature profiles as a function of the polar angle for 
$B_p=10^{15}$ G and three different combinations of the $T_b$--distribution
and MF geometry.
The solid lines correspond to the numerical results and the dashed lines to 
the analytic fits from Eqs. (\ref{fitsnew}) and (\ref{fit3}).
}
\label{surface}
\end{figure} 
Concerning 2D models,
solid lines in Fig. \ref{fig1} correspond to our
2D transport results with a dipolar field. 
The quoted value of the MF corresponds to the strength at the
magnetic pole.  
From our results, we confirm that the effect of ion/phonon transport in the envelope
is to reduce the large anisotropy obtained in previous magnetized models. 
In addition,
by performing 2D heat transport calculations through the magnetized envelope, 
we take into account the meridional heat fluxes driven by the meridional temperature 
gradients between pole and equator, and we find that the anisotropy is further
reduced. This effect is more relevant for high fields, resulting in an equatorial
temperature about a factor 3 lower than the polar temperature, in contrast with
the 2-3 orders of magnitude obtained in previous models.
For practical purposes, we have made fits of our results keeping a similar
functional form to Eq. \ref{fitsold}, but only changing the form of 
$\mathcal{X}_\parallel$ and $\mathcal{X}_\perp$ as follows:
\ba
\mathcal{X}_\parallel(B,T_{\rm b}) &=& 1 + 0.05 B_{12}^{0.25} T_{b,9}^{0.240}
\nonumber \\
\mathcal{X}_\perp(B,T_{\rm b}) &=& \frac{\sqrt{1 + 0.07 B_{12} (0.03+T_{b,9})^{-0.559}}}
{\left[1 + 0.9 B_{12}/ (0.03+T_{b,9})\right]^{0.4}}
\label{fitsnew}
\ea
We find that this formula reproduces very well the correct numerical results for 
the $T_{\rm b}-T_{\rm s}$ relation for a dipolar field geometry.
The ratio between polar and equatorial temperature is significantly
lower than in the case of 1-D plane-parallel envelope models.
We also confirm that, if neutrino emission at densities 
$\rho> 10^{10}$ g/cm$^3$ is taken into account,
there is a MF-dependent upper limit to the effective surface temperature
\citep{Potekhin2007}, but our maximum temperatures for MF tangential to the surface
are considerably higher than those obtained before.
We find that the upper limit consistent with our numerical results
is well fitted by the following expressions
\ba
T_{s,\parallel}^{\rm max} = \frac{3.6 \times 10^6 {\rm K}}
{1 + 0.02 \log{B_{12}}}
\nonumber \\
T_{s,\perp}^{\rm max} = \frac{2.8 \times 10^6 {\rm K}}
{1 + 0.6 \log{B_{12}}}
\ea

It must be stressed that these upper limits are obtained by the inclusion of the
outer crust ($3\times 10^{11} > \rho> 10^{10}$ g/cm$^3$) in the static envelope calculations that
fix the internal boundary condition. If the numerical
grid of the cooling code extends to densities lower than $10^{10}$ g/cm$^3$
this effect is consistently incorporated and there is no need to include these
corrections in the $T_{\rm b}-T_{\rm s}$ relation.
Note that we neglect the possibility of Joule heating within the envelope because,
due to the very large magnetic diffusivity, initial electric currents are quickly dissipated
and may only be important at the very beginning of a NSs life.
Obviously these results may vary for other field geometries. We have checked, however,
that our improved fits (\ref{fitsnew}) combined with Eq. (\ref{fit3}) are a good approximation
(within $5\%$) to the numerical result in a number of cases.
In Fig. (\ref{surface}) we compare the numerical results with the analytical fits
for three different models. In the top panel the MF geometry 
is purely dipolar and the temperature distribution at the inner boundary
is fixed to $T_b=10^7 (\cos^2\theta + 0.2 \sin^2\theta)$ K. The middle panel shows
the results for a model with the same MF configuration but a temperature
distribution at the base of the envelope given by 
$T_b=10^7 (1 + 3 \sin^2\theta)$ K. The bottom panel corresponds to a model
with the same $T_b$ distribution as the middle panel, but with a purely 
quadrupolar MF geometry. In all cases $B_p=10^{15}$ G, so that the combination
of high field and low temperature demonstrates the validity of the fit for the
most extreme cases.

\begin{figure}
\centering
\includegraphics[width=0.45\textwidth]{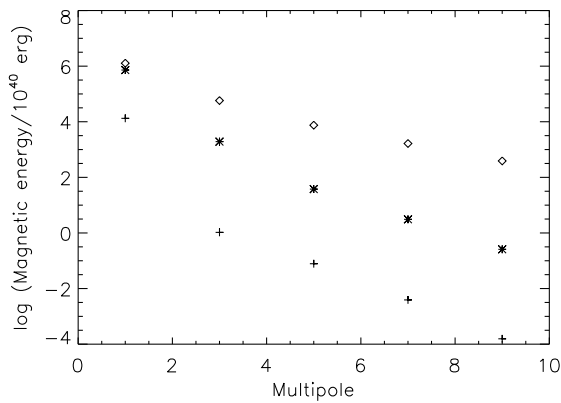}
\caption{Power spectrum at  $t=5\times10^5$ years for the 
models corresponding to 
purely poloidal fields with initial values of 
$B_p=  10^{13}$ (crosses), $10^{14}$ (stars), and $10^{15}$ G (diamonds).
}
\label{spect}
\end{figure}

\section{Results}

\begin{figure*}[th]
\centering
\includegraphics[width=0.95\textwidth]{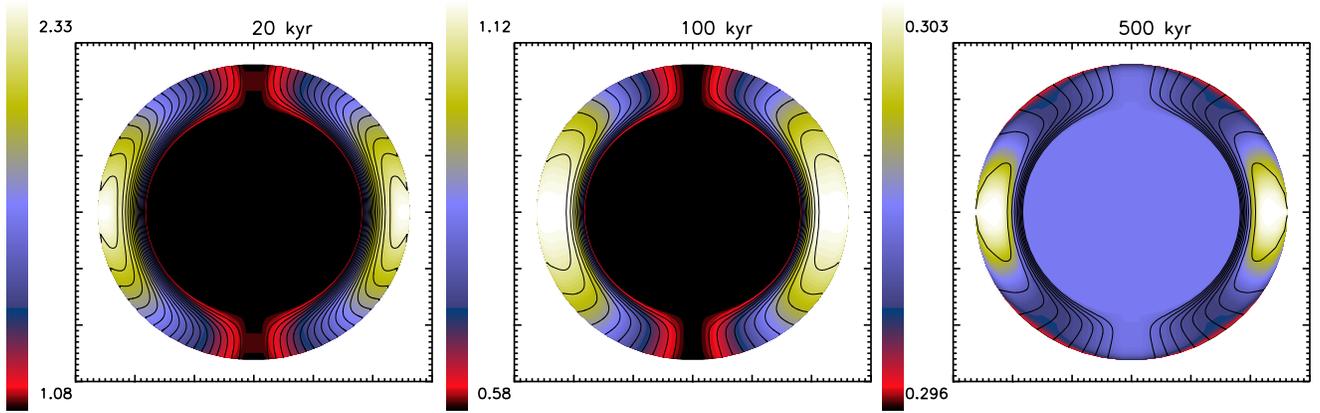}
\caption{ Temperature distribution in the crust of a NS (i.e. up to the bottom of 
the envelope) at different
ages. The initial MF is purely poloidal with
($B_p=10^{14}$ G). The field lines are also shown.
The numbers in the color-scale on the left of each figure indicate
the maximum and minimum values of the temperature (in units of $10^8$ K) at each age.
The crustal shell has been stretched a factor of 4 for clarity.
}
\label{figcolor}
\end{figure*}

We present the results now of our numerical simulations of the magneto--thermal
evolution of NSs and its dependence on the initial MF structure and strength.
We restrict this presentation to a selection of the initial models described
in \cite{Aguilera2008b}. Our model A at $t=0$ consists of a dipolar
($n=1$) poloidal field according to Eq. (13) of \cite{Aguilera2008b}, parameterized
by the value of the radial component at the magnetic pole, $B_p$. In model B
we add to the initial dipolar field a quadrupolar ($n=2$) toroidal
component obeying Eq. (11) of \cite{Aguilera2008b}, with a
maximum value of $\approx 50 B_p$. This latter model represents the case
of a strong toroidal component confined in the crust.
We recall that we focus in this paper on the purely diffusive case, i.e., the Hall term
in the induction equation is neglected. This term needs a separate specific numerical
treatment that is out of the current capabilities of our code and will be discussed
in later work. Its influence can be important either at very early times or,
as we point out at the end of the section, at late times during the photon cooling era.
For most of the first $10^5-10^6$ years of a NS life we do not expect qualitative
changes from our present results.
 
To begin our discussion, we present in Fig. \ref{spect} the power spectra
for model A with three initial fields of 
$B_p=10^{13}$, $10^{14}$, and $10^{15}$ G at the age of $t=5\times10^5$ years.
We see how the coupling between different modes due to the angular
dependence of $\eta$ fills the shorter wavelength modes (initially only 
the dipolar poloidal component $n=1$ is present). Only odd multipoles
are present because the initial model is symmetric with respect to the
equatorial plane.
At this age, the cascade has filled out all large wavenumber modes and is
saturated following approximately an $n^{-4}$ power law for the cases with
large initial fields. 
This result shows that, even if the influence of the non-linear Hall 
term is negligible, the thermo--magnetic coupled evolution results in a complex field geometry
that can not be described by a single mode.
Although the dominant mode is still the dipole, the distribution of part of the energy
in higher other modes leads to a slightly faster dissipation.

In Fig. \ref{figcolor} we show the crustal temperature distribution of a NS at different
ages ($t=10, 100$ and 500 kyr), together with the poloidal field lines. 
The tendency of the surfaces of constant temperature to be aligned with the magnetic 
field lines discussed in \cite{Azorin2006a,Geppert2004}
is clearly visible, but there is an important difference with respect to previous works.
The consistent inclusion of the Joule heating source results in a greater energy deposition in
the region where currents are more intense. In this particular model this happens at 
low latitudes,
since the currents that maintain a poloidal field are toroidal. Thus more heat is
released close to the equator. While the polar regions are always in thermal equilibrium with the
core, the equator is insulated due to the reduced thermal conductivity across magnetic 
field lines.
As a consequence, the heat released at early times in the equatorial region cannot flow across
field lines into the core, where it would be rapidly lost by neutrino emission. 
It is only allowed to flow towards the surface along field lines.  
This modifies the traditionally accepted temperature distribution consisting 
of hot poles and a cooler equator 
(insulated from the warmer core). Instead we find that the equatorial region 
of the outer crust is actually warmer than the poles. 
In models with weak MFs, when the effect of Joule heating becomes
less effective, the situation is inverted and the 
thermal distribution with hot polar caps is recovered. 
Notice the large difference 
between polar and equatorial temperatures, up to a factor of 2
at early times (the numbers next to the color scale indicate the maximum and 
minimum temperature at each evolutionary time), in contrast with the nearly
isothermal crusts at late times.

In non--magnetized NSs the heat conduction between
the crust and the core is able to counteract the local neutrino emission
processes and produce an isothermal NS after the thermal relaxation 
stage, which lasts 10-100 years. This interchange of energy between core
and crust is still important in magnetized neutron stars, but the presence
of a strong tangential field in the equatorial region of the inner crust has the effect
of increasing the thermal relaxation time in several orders of magnitude,
so that the equatorial region in Fig. \ref{figcolor} becomes effectively
decoupled from the core.

\begin{figure}[th]
\centering
\includegraphics[width=0.45\textwidth]{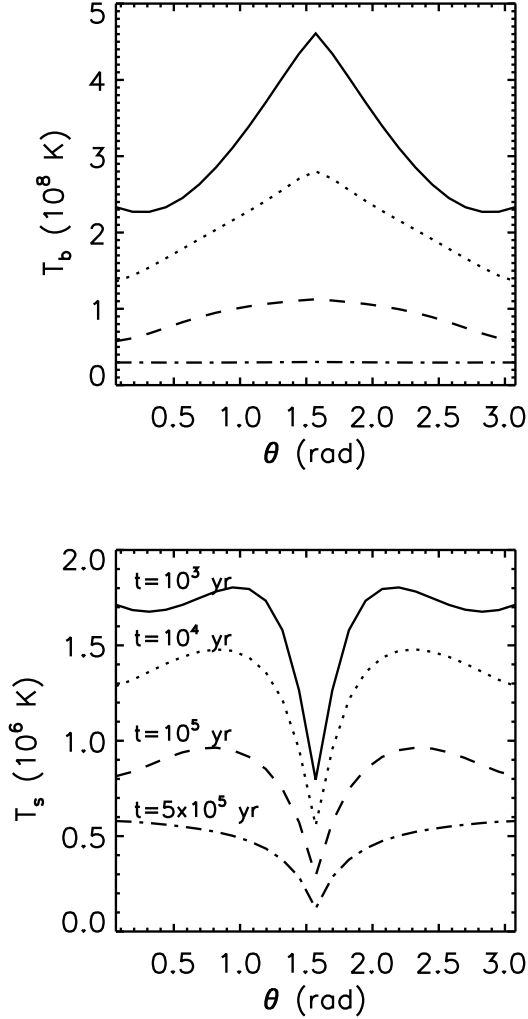}
\caption{Temperature profiles at the base of the envelope
($T_b$) and at the surface ($T_s$) at different evolutionary times.
The initial MF is purely poloidal with $B_p=10^{14}$ G.
}
\label{tbts}
\end{figure}
We must remind that the anisotropy in $T_{\rm b}$ (at the bottom of the envelope), 
is not necessarily the same as that of $T_{\rm s}$. The blanketing effect of the 
envelope, and/or atmospheric effects should be taken into account before a 
comparison with observations. In Fig. (\ref{tbts}) we plot the evolution of both,
the temperature at the bottom of the envelope (top panel) and the surface
temperature (bottom panel) as a function of the polar angle.

\begin{figure}[th]
\centering
\includegraphics[width=0.45\textwidth]{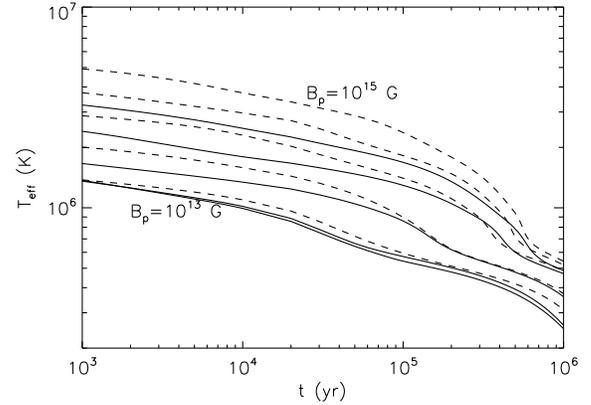}
\caption{
Cooling curves. Effective temperature as a function of age
for different initial field strengths (from bottom to top
$B_p=  10^{13}, 3 \times 10^{13}, 10^{14},  3 \times 10^{14}$, and $10^{15}$ G).
The solid lines correspond to Model A and the dashed lines to model B.
}
\label{fig4}
\end{figure} 

In Fig.~\ref{fig4}, we show a sample of cooling curves (effective temperature versus
true age) for models A (solid lines) and B (dashed lines) but varying the initial strength 
of the field. For high fields the effect of Joule heating
is visible from the very beginning of the evolution. The effective temperature of a young,
$t=10^3$ yr magnetar with $B_p=10^{15}$ G is higher by a factor of $2$ than that of a
NS with a standard $B_p=10^{13}$ G, and it is kept nearly
constant for a much longer time. The effect is further enhanced
in model B due to the extra energy stored in the toroidal field.
In spite of some quantitative differences, our improved simulations confirm the
qualitative results described in \cite{Aguilera2008a}, where a phenomenological
MF decay law has been adopted. The most important difference is the somewhat higher
temperature reached in these models, when compared to \cite{Aguilera2008a}.
The difference stems from the particular location of the heat dissipated
by the Joule effect in the regions where currents are intense, as opposed to
the homogeneous distribution of energy in the formerly adopted phenomenological
model. 

\begin{figure}[th]
\centering
\includegraphics[width=0.45\textwidth]{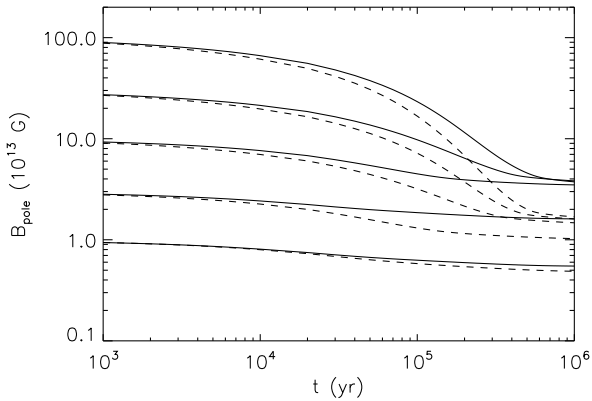}
\caption{Evolution of the MF strength at the pole $B_p$ during the first million
years of a NS for several initial values.
The solid lines correspond to Model A and the dashed lines to model B.
}
\label{fig5}
\end{figure}

In Fig. \ref{fig5} we show the value of the MF at the pole as a function of time
for models A (solid lines) and B (dashed lines) with different initial field strengths.
Naively, we can divide the models in two groups: strong initial 
field $B>5\times 10^{13}$ G and weak initial field $B<5\times 10^{13}$ G.
The plot shows that models with strong initial fields are subject to a faster decay 
than those with weaker fields.
Models with weak initial fields are subject to decay in a more or less similar manner. The
field typically decays in about a factor of two on a timescale of few $10^6$ years, and then
remains nearly constant, due to the increase of the magnetic diffusion timescale as the
star cools down. 

Models with initially large fields behave in a different way. Since the magnetic energy 
stored in the crust is now large enough to significantly affect the thermal evolution 
when it is steadily released by Joule heating, it results in a higher average temperature 
of the crust. But this process
has a back-reaction: the higher temperatures imply larger resistivities and therefore faster
decay. Interestingly, at $t>10^6$ yr, all models with large poloidal initial fields seem 
to converge to an asymptotic fiducial value of 
$B_{\rm asymp} \approx 4 \times 10^{13}$ G, while
all models with strong toroidal fields converge to a lower value of 
$B_{\rm asymp} \approx 1.5 \times 10^{13}$ G, because of the higher temperatures
reached in average during the evolution.

Based on these results we predict that all
sufficiently old NSs born with $B_p$ above a critical value evolve towards
similar field strengths, while those born with lower fields show a MF distribution
at late times similar to the initial distribution but shifted in about a factor of 2
to lower values.  This critical initial field is approximately 
$B_{\rm crit} \approx 5 \times 10^{13}$ G.
It also delimits
the minimum field strength that can actually influence the thermal evolution of the NS
by MF diffusion. 
Once the MF is dissipated below that value, its influence on the later evolution is reduced
and the subsequent evolution proceeds in a similar way in all cases.
Our models predict that no old magnetar can be found, and that,  at ages $t>5\times 10^5$ yr,
the typical MF for all NSs born as magnetars must be similar ($B \approx B_{\rm asymp}$).
Up to know, observational data are not in contradiction with this fact.

To conclude this section we turn now to discuss the evolution of the
dimensionless magnetization parameter $\omega_B \tau$, where $\omega_B$
is the electron gyro--frequency and $\tau$ is the electron relaxation time.
The Hall induction equation (\ref{Hallind}) can also be written as 
\begin{equation}
\frac{\partial \vec{B}}{\partial t}= - \frac{c^2}{4\pi} \vec{\nabla}\times\left(
\frac{1}{\sigma_\parallel} \left\{ \vec{\nabla}\times (e^{\nu}\vec{B})+
{\omega_B \tau}\left[\vec{\nabla}\times(e^{\nu}\vec{B})\right] \times \vec{e_B}~
\right\} \right)~,
\end{equation}
where $\vec{e_B}$ is the unit vector in the direction of $B$.

\begin{figure}[th]
\centering
\includegraphics[width=0.45\textwidth]{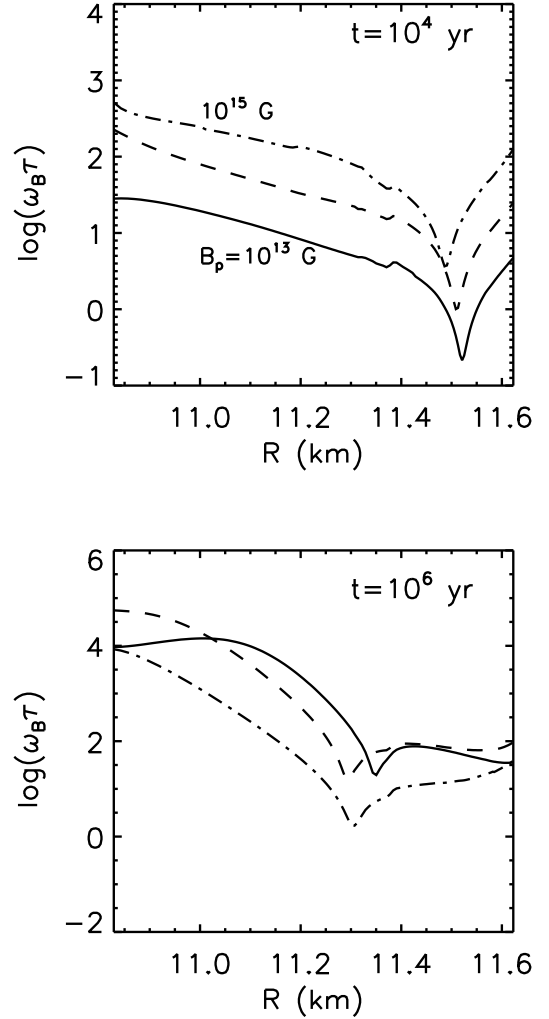}
\caption{Radial equatorial profiles of the magnetization parameter $\omega_B \tau$ for
model A and three different initial values of $B_p$: $10^{13}$ G (solid lines), 
$10^{14}$ G (dashed lines), and $10^{15}$ G (dotted lines).
The upper panel corresponds to $t=10^4$ yr and the lower panel to $t=10^6$ years.
}
\label{fig6}
\end{figure}

In this form, the interpretation of $\omega_B \tau$ is straightforward: when the
magnetization parameter exceeds unity, the Hall drift term dominates.
The effect of the Hall drift term at early times has been discussed in detail in
\citep{PonsGeppert2007}. As a rule of thumb, when $\omega_B \tau > 10^3$ the effect
of the Hall drift significantly changes the evolution, while more moderate values 
lead to a somewhat faster dissipation due to reorganization of the field in smaller
scales.  Due to numerical limitations, in this work we have restricted ourselves
to the purely diffusive case, but it is worth to look at the evolution
of this parameter as shown in Fig. \ref{fig6}. Here we 
show radial profiles of the magnetization parameter in the crust
for model A and three different initial values of $B_p$.
The upper panel corresponds to $t=10^4$ yr and the lower panel to $t=10^6$ years.
At early times, the scaling of $\omega_B \tau$ with $B_p$ is visible
and only in magnetars one must expect large values in the inner crust.
However, the back-reaction of the field evolution on the temperature has some
interesting implication. At late times, the temperature of low field NSs is lower
than that of highly magnetized NS, so that the temperature dependence of the
electron relaxation time overcomes the effect of the MF and it turns out
that the former magnetars are {\it less magnetized} than NSs born with moderate
fields. In addition, at ages of $\approx 10^6$ yrs the temperature is low enough
to ensure that the magnetization parameter is very large for all models studied,
specially in the inner crust (where the conductivity is larger). This opens
more questions about the late evolution of NSs that is likely to be dominated
by the non--linear Hall term in most NSs. In particular, the possibility of the 
Hall instability \citep{RG02}, or implications on the evolution of the braking index 
are worth to be explored.

\section{Conclusions}

We have performed consistent 2D simulations of the coupled magneto-thermal
evolution of NSs for the first time, including realistic microphysical input and 
general relativistic corrections. By properly taking into account the interplay of 
the MF and temperature evolution in cooling simulations we have 
found that their mutual influence is important
and affects the outcome when the initial field strength is of the order
or larger than a critical value of $B_{\rm crit} = 5 \times 10^{13}$ G.
It has effects on both sides: the temperature and the field strength.

The average effective temperature of a NS born with $B_p <  B_{\rm crit}$
is barely affected, while those born as magnetars are subject to significant
heating by the dissipation of currents in the crust. In addition, since
heating is locally important in the regions where currents are more intense,
the surface temperature distribution can be very different depending on
the field geometry. While the pole (or other regions in which the field
lines are nearly radial through the crust) is in thermal equilibrium with the
core and has its same temperature, regions in which the field lines are tangential
remain essentially thermally isolated. This can produce cooler areas if no
heating process is considered or, conversely, hotter regions if heating is
important. This is the case of poloidal fields, in which currents are located
near the equatorial region that remains warmer than the rest of
the crust during a long time.
The effective temperature of
models with strong internal toroidal components are systematically higher than that
of models with purely poloidal fields, due to the additional energy reservoir stored
in the toroidal field that is gradually released as the field dissipates.

In the models with stronger fields,
as a result of the average higher temperatures,
the crustal electrical resistivity is enhanced and magnetic diffusion proceeds faster
during the first $10^5-10^6$ years of a NS's life. As a consequence, 
all NSs born with fields larger than a critical value ($> 5 \times 10^{13} G$)
reach similar field strengths ($\approx 2-3 \times 10^{13} G$) at late times,
irrespectively of the initial strength. 
After $10^6$ years the temperature is so low that the magnetic 
diffusion timescale becomes longer than the typical ages of radio--pulsars, 
resulting in apparently no dissipation of the field in old NSs. We 
confirm the strong correlation between the MF and the surface temperature
of relatively young NSs discussed in preliminary works. 
Notice that if the MF of magnetars is caused by superconducting currents in the core
(as opposed to crustal currents) the longer diffusion timescale in the core 
would allow magnetars to live much longer with their original large fields. 
Thus, observations of magnetars can help to discern between models with currents 
located in the crust or in the core.

It should be mentioned that magnetic field evolution by ambipolar diffusion in
the core may produce qualitatively similar effects to those obtained in this work:
keeping the temperature high while the field is strong and stopping field decay
when the temperature drops \citep[see e.g.][]{Reis2008}. 
In this work we have not considered the MF evolution in the core
because ambipolar diffusion is usually considered under the
assumption that the core is in a non--superfluid state and therefore is
important during the very early stages of evolution. We are more interested
in the long-term evolution, after the temperature rapidly drops below the critical
temperature for nucleon superfluidity. In order to quantify
the relative importance of both effects one would need to consider
ambipolar diffusion in a superconducting fluid coupled 
to the dissipation of crustal currents studied here.

The detection of magnetars with {\it true} ages (the spin-down age can
be seriously overestimated) $t \approx 10^6$ yr, or the detection of a young highly magnetized
NS with $T<10^6$ K would be a serious challenge for crustal field
models. At present our results are in agreement with the known population of high field
NSs and magnetars and support the idea of the existence of a strong crustal MF
component in magnetars.

Given the complexity of the feed-back between temperature and MF,
it seems necessary to extend this work in two main lines that can shed new light
on our knowledge of the cooling theory of NS: engaging 3D simulations
and including the Hall term in the induction equation. The complex geometry
that may arise in a realistic case with hot spots and irregular fields is certainly
not treatable with our present code and needs further investigation. Similarly,
the influence of the Hall term at late times (when the temperature is so low
that the magnetic diffusivity is negligible)
or a consistent treatment of ambipolar diffusion in a superconducting 
fluid coupled to the dissipation of crustal currents,
can produce new interesting effects and are issues worth to explore in future works.

\begin{acknowledgements}
We thank D.N. Aguilera for valuable comments and providing updated conductivity
routines used in the simulations, and Andreas Reisenegger for a critical reading
and his constructive and helpful comments. This research has been supported by the 
Spanish MEC grant AYA 2007-67626-C03-02 and the Research Network Program {\it Compstar}
funded by the ESF. U. Geppert thanks the University of
Alicante for support under its visitors program.
\end{acknowledgements}

\bibliography{cooling} 

\end{document}